**Spatially Uniform and Defect-Tolerant Plasmonic Responses in 3D printed Gold Nanoparticle Assemblies**

*Vasanthan Devaraj, Sunghyun Kwak, Hyeongjip Kim, Sang-Keun Sung, Jong-Min Lee*, Thomas Zentgraf * and Won-Geun Kim**

Vasanthan Devaraj, Thomas Zentgraf
Department of Physics, Paderborn University, 33098 Paderborn, Germany
Institute for Photonic Quantum Systems (PhoQS), Paderborn University, 33098 Paderborn, Germany
E-mail: thomas.zentgraf@uni-paderborn.de

Sunghyun Kwak, Hyeongjip Kim, Won-Geun Kim
Department of Optical Engineering, Kumoh National Institute of Technology, Gumi 39177, Republic of Korea
E-mail: wgkim@kumoh.ac.kr

Sang-Keun Sung
Carbon Neutrality & Bio Research Center, Gumi Electronics & Information Technology Research Institute, Gumi 39253, Republic of Korea

Jong-Min Lee
School of Semiconductor Display Technology, Hallym University, Chuncheon 24252, Republic of Korea
School of Nano Convergence Technology & Nano Convergence Technology Center, Hallym University, Chuncheon 24252, Republic of Korea
E-mail: jmlee@hallym.ac.kr

Won-Geun Kim
Department of Physics, Kumoh National Institute of Technology, Gumi 39177, Republic of Korea

Abstract

Three-dimensional (3D) assemblies of gold nanoparticles (AuNPs) offer a rich platform for plasmonic coupling and near-field engineering, yet their optical behavior is often complex due to structural disorder and fabrication-induced variability. Here, we present a systematic optical investigation of large-scale 3D AuNP assemblies fabricated via meniscus-guided assembly, focusing on the reproducibility and spatial uniformity. Spatially-resolved dark-field scattering measurements reveal that high-aspect-ratio AuNP pillars exhibit uniform scattering spectra along their height and across different pillars, despite variations in geometry and structure. Electromagnetic simulations suggest that this robustness arises from an ensemble-averaged plasmonic response governed by many local coupling regions within a finite plasmon delocalization length. Simulated near-field and surface charge distributions suggest that the broad ensemble response remains spatially distributed under representative structural perturbations, consistent with volumetric averaging. Building on this robust platform, we introduce compositional modulation through a core–satellite architecture by incorporating smaller AuNPs. This yields a composition-dependent spectral redistribution, including an additional long-wavelength spectral feature in the core–satellite assemblies. Wavelength-dependent surface-enhanced Raman scattering measurements reveal contrasting responses under 633 and 785 nm excitation, reflecting redistribution of local plasmonic coupling pathways. These results establish design principles for robust 3D plasmonic nanoparticle assemblies with ensemble-averaged and composition-tunable optical responses.

Vasanthan Devaraj and Sunghyun Kwak equally contributed to this work.

## 1. **Introduction**

Small gold nanoparticle (AuNP) clusters and oligomers exhibit rich optical properties arising from coupled localized surface plasmon resonances (LSPRs).[1-3] When light drives the coherent oscillation of electrons in closely spaced AuNPs, an intense local-field enhancement originates from the nanogaps or "hot-spots" [4,5], and the coupled plasmon modes of the cluster can differ markedly from those of isolated particles.[6,7] Such plasmonic mode coupling in AuNP assemblies gives rise to geometry-dependent phenomena including strong local field enhancement.[8-11] These capabilities make AuNP clusters promising for a broad range of nanophotonic applications - from molecular sensors [12,13] and surface-enhanced spectroscopies to metasurface devices [14,15] - where the manipulation of light at subwavelength scale is crucial. The specific LSPR characteristics of a cluster (peak wavelength [16], intensity [17], line shape [18]) are highly sensitive to its architecture (particle size [16,17], number [19], arrangement [20], and inter-particle gaps [21]) and environment [16,22], offering a versatile design space to tailor optical responses via nanostructure engineering. Realizing complex three-dimensional (3D) AuNP cluster architectures, however, poses significant fabrication challenges. Traditional top–down nanofabrication methods such as electron-beam or ion-beam lithography can create precise NP patterns, but they are typically limited to planar (2D) layouts and suffer from low throughput and high cost.[23,24] Stacking multiple lithography layers to achieve 3D structures is cumbersome, and structural reproducibility can be an issue even for 2D patterns.[25] Bottom–up self-assembly routes (e.g. DNA-origami linking of colloids [26,27] or template-assisted convective assembly [28,29]) offer a pathway to build 3D AuNP clusters in a more scalable fashion. Yet, these chemical assembly approaches rely on thermodynamic or surface-specific driving forces, which restrict the attainable geometries and make it difficult to precisely control particle placement – especially for heterogeneous clusters containing different NP types.[30,31] For instance, incorporating a second component (like a quantum dot or a different metallic NP) into a colloidal AuNP cluster often requires elaborate surface functionalization (linker molecules, DNA scaffolds, etc.) to guide the assembly. Consequently, there is a need for fabrication techniques that can reliably create freestanding 3D AuNP assemblies with controlled macroscopic geometry, tunable nanoscale composition, and reproducible ensemble optical properties. Meniscus-guided 3D nanoprinting has recently emerged as a promising strategy to meet these needs.[32-34] In this approach - often implemented via a glass micropipette - a tiny meniscus of colloidal ink is formed at the nozzle tip and used to deposit NPs in a directed manner.[32,35] As the solvent evaporates from the femtoliter-scale meniscus, AuNPs are continuously drawn to the contact point and immobilized, allowing the structure to grow under the guidance of the moving tip.[36] By translating the micropipette in three dimensions, one can write complex freeform NP assemblies and even extended architectures like pillars, helices, or zigzag wires, effectively overcoming the planar constraint of lithography. This meniscus-guided printing method enables highly localized assembly of AuNPs into closely packed, freestanding 3D nanoparticle architectures with nanoscale interparticle separations.[4,5,21,37] Importantly, it also provides exceptional materials flexibility: multiscale and heterogeneous clusters can be fabricated in one step by simply using an ink that contains a mixture of NP types.[38] Meniscus-guided 3D nanoprinting thus offers unprecedented control over 3D cluster geometry and composition, opening up new design possibilities for plasmonic structures.[35] Despite this progress, a fundamental understanding of the optical behaviors of large 3D-printed AuNP assemblies remains limited.[35,38] To date, most studies have focused on device-level demonstrations (improving signal outputs or showcasing new functionality) rather than systematically interrogating how the plasmonic modes in these architectures behave in space and spectrum.[35,39-41] For instance, the uniformity of the plasmonic scattering response across different height measurement points from a tall pillar, or how sensitive is the near-field enhancement distribution in presence of structural imperfections (example: a missing NP, disordered nanogaps or a slight variation in NP packing density). Core–satellite plasmonic structures have also been predicted to support radiative and less-radiative coupled responses depending on the relative arrangement of cores and satellites.[42-44] However, experimentally resolving and interpreting such coupled responses in large, disordered 3D nanoparticle assemblies remains challenging. In short, the spectral and spatial characteristics of large plasmonic 3D nanoparticle assemblies fabricated by meniscus-guided printing have not yet been comprehensively characterized, especially with respect to height-dependent LSPR uniformity [45], defect-induced changes in near-field hotspots [46], and the excitation of Fano-like dark resonances in deliberately engineered cluster geometries.[47,48] Here, we address these knowledge gaps by conducting an in-depth optical investigation of large-scale 3D AuNP assemblies produced via meniscus-guided 3D nanoprinting. In typical small-N oligomer systems, such as dimers, trimers, and heptamers, optical properties are highly sensitive to nanometer-scale variations in the interparticle gap or to missing particles that break symmetry.[49,50] Even minor structural imperfections can dramatically shift resonance modes or reduce near-field enhancement, limiting their robustness and scalability.[51] In contrast, the printed structures investigated here contain a much larger number of nanoparticles and are more appropriately viewed as large, densely packed 3D nanoparticle assemblies rather than discrete oligomeric clusters. In these assemblies, the optically probed response does not require coherent delocalization over the entire pillar. Instead, it arises from many local coupling regions whose contributions are averaged within the finite plasmon delocalization length and the experimental sampling volume. This distinction explains why the measured scattering spectra are relatively insensitive to macroscopic pillar dimensions while remaining representative of the local nanoscale packing statistics. This ensemble-averaged response provides the physical basis for the spatial uniformity and reproducibility discussed throughout this work. This distinction also separates the present system from conventional nanoparticle chain geometries, where the optical response is often governed by one-dimensional coupling along a defined particle sequence and by polarization-dependent longitudinal mode hybridization. Here, the relevant response arises from a three-dimensional distribution of many local coupling regions, leading to an ensemble-averaged scattering response rather than a chain-like plasmon mode. We combine spatially resolved dark-field scattering measurements and full-field electromagnetic simulations to unravel the physical origins of the optical responses of nanoparticle assemblies. In particular, we examine how the scattering spectra distributions are maintained or altered from the base to the top of high-aspect-ratio nanoparticle pillars, how defined defects or variations in the cluster structure influence local field enhancements, and how core–satellite compositional modulation redistributes local plasmonic coupling and wavelength-dependent near-field responses. By combined experimental and theoretical studies, regarding the reproducibility and robustness of optical modes, we aim to derive the design principles for plasmonic NP architectures that are tolerant to structural imperfections while retaining desired optical functionalities.

## 2. **Results and Discussion**

In this process, a glass micropipette with a small aperture is used to print an aqueous dispersion of AuNPs onto a substrate while maintaining a confined liquid meniscus at the tip (**Figure 1a**). As the micropipette is translated relative to the substrate, the meniscus serves as a localized assembly region for nanoparticle deposition. AuNPs dispersed in water are continuously transported toward the three-phase contact region, where evaporation-driven accumulation leads to the formation of a freestanding nanoparticle assembly. Rather than imposing long-range order through templates or chemical linkers, this approach relies on the confinement and guidance of a microscale meniscus to enable a localized, continuous assembly. As a result, the internal structure of the pillar is intrinsically disordered, yet densely packed, reflecting a statistical assembly of NPs within a spatially constrained volume. A defining feature of this meniscus-guided assembly process is the steady and continuous supply of NPs to the growth front once the meniscus geometry is stabilized. Under such conditions, NPs incorporated at different stages of the growth experience similar local environments, leading to comparable packing statistics along the vertical direction of the structure. Although individual NP arrangements vary locally, the absence of abrupt changes in the assembly conditions suggests that the pillar maintains broadly comparable packing characteristics along the growth direction. **Figure 1b** further illustrates the compositional versatility of the assembly process by illustrating the formation of core–satellite-type AuNP assemblies using a hybrid NP ink. Without altering the underlying assembly mechanism, the introduction of smaller AuNPs leads to multiscale clusters in which satellites decorate the surfaces and interstitial regions of larger cores. This capability highlights that structural complexity and plasmonic coupling pathways can be modulated primarily through NP composition, while preserving the statistical nature of the assembly. The optical micrographs in Figure 1c provide direct experimental evidence of the meniscus-guided assembly process, capturing representative stages of (i) contact, (ii) growth, and (iii) detachment during the pillar formation. In these experiments, an aqueous dispersion of polyvinylpyrrolidone (PVP)-coated AuNPs (80 nm diameter) was supplied through a glass micropipette with an aperture of approximately 5 μm. Assembly was carried out under ambient conditions (relative humidity ~ 20%), where solvent evaporation from the confined meniscus continuously drove NP accumulation at the growth front.

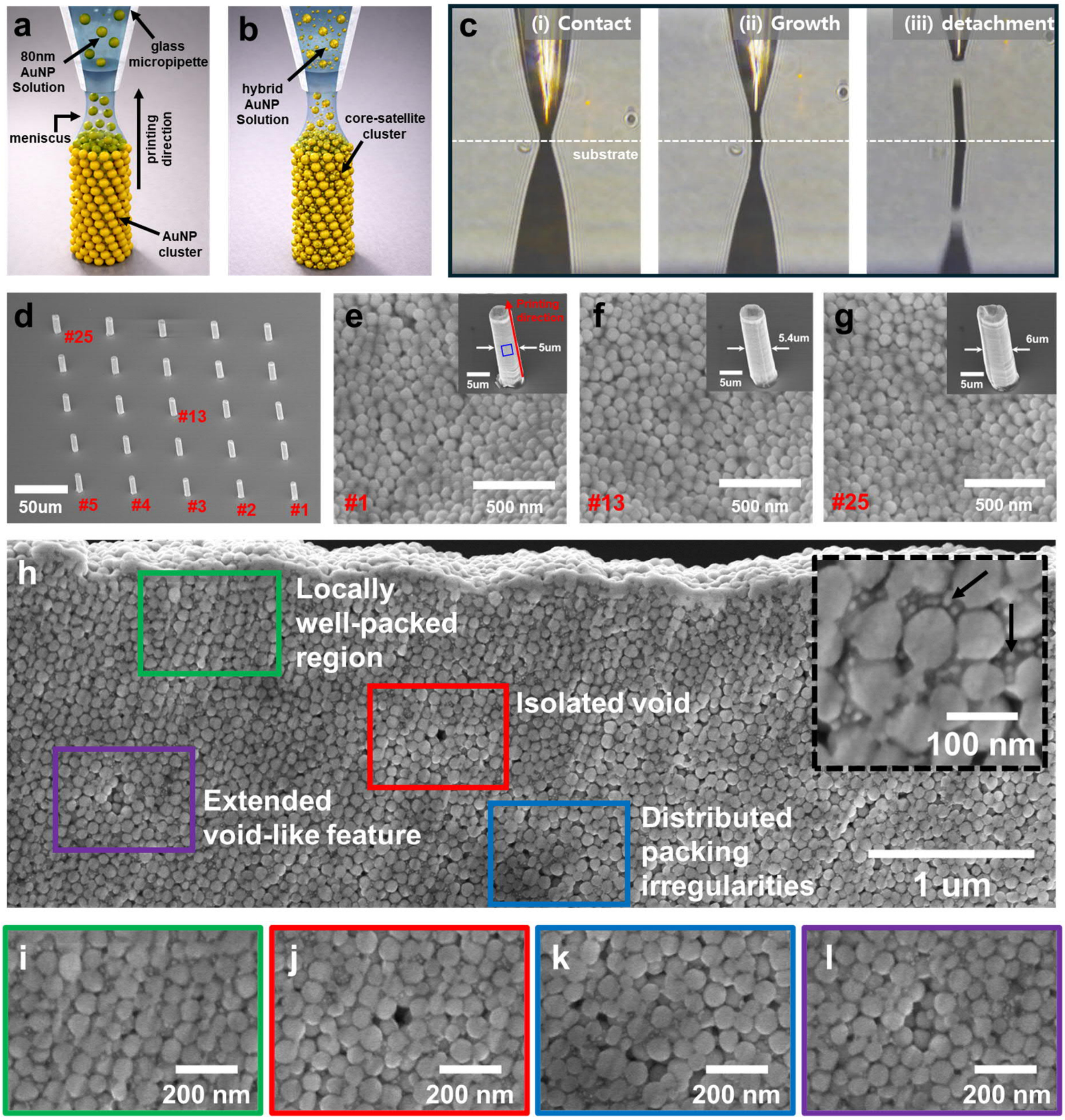

**Figure 1 | (a)** Schematic illustration of the meniscus-guided 3D printing of AuNPs. **(b)** Schematic illustration of core–satellite AuNP assembly formation using a hybrid nanoparticle ink containing AuNPs of different sizes. **(c)** Optical micrographs showing representative stages of the assembly process: (i) initial contact between the glass micropipette and the substrate, (ii) vertical growth of an AuNP pillar, and (iii) detachment after completion of assembly. **(d)** SEM image of a 5 × 5 array of freestanding AuNP assembly pillars fabricated by meniscus-guided assembly. **(e–g)** Magnified side-view SEM images of representative printed AuNP assembly pillars. The inset in (e) indicates the printing direction (red solid arrow) and the representative magnified region selected from the middle-height region of the pillar (blue solid box). **(h)** Large-area FIB cross-sectional SEM image of a printed core–satellite AuNP assembly pillar containing 80 nm AuNPs and intentionally incorporated 20 nm satellite AuNPs. The inset highlights the smaller satellite AuNPs within the cross-section. **(i–l)** Magnified cross-

sectional SEM images showing representative local packing characteristics, including (i) relatively well-packed regions, (j) isolated voids, (k) distributed packing irregularities, and (l) extended void-like regions.

By translating the micropipette upward at a constant speed of 300 nm/sec, freestanding AuNP pillars with typical heights of ~20 μm were reproducibly formed. The sequential images in Figure 1c reveal a continuous and stable growth process without visible interruptions or morphological transitions, followed by a well-defined detachment step. This experimental observation supports the premise that the assembly proceeds under quasi-steady conditions, providing a structurally consistent basis for subsequent analyses of spatially resolved optical properties. **Figure 1d–g** presents the structural morphology and reproducibility of the printed 3D AuNP assembly pillars fabricated by meniscus-guided assembly. The SEM image of a 5 × 5 pillar array (**Fig. 1d**) shows the formation of freestanding pillar arrays, while the magnified side-view SEM images (**Fig. 1e–g**) reveal dense and locally disordered nanoparticle packing within representative pillar bodies. The optical uniformity and the influence of macroscopic pillar-to-pillar diameter variations are evaluated separately using dark-field scattering measurements in **Figure 2**. The large-area cross-sectional SEM image in **Figure 1h** was obtained from a printed core–satellite AuNP assembly prepared using a hybrid ink containing 80 nm and 20 nm AuNPs, as schematically illustrated in **Figure 1b**. The smaller particles visible in the magnified cross-section SEM image (**inset of Fig.1h**) are intentionally incorporated 20 nm satellite AuNPs, rather than unintended impurities or debris. This cross-section provides representative structural evidence that smaller AuNPs can be incorporated into the internal interparticle regions of the printed 3D assembly, rather than being limited to the outer surface. At the same time, the image reveals that the printed assembly is dense but locally disordered, containing locally well-packed regions (**Fig. 1i**), isolated voids (**Fig. 1j**), distributed packing irregularities (**Fig. 1k**), and extended void-like features (**Fig. 1l**). Such local disorder can arise from multiple factors, including the finite size distribution of the colloidal AuNPs, possible particle shape variations, stochastic packing during meniscus-guided assembly, and local rearrangement during solvent evaporation. We note that these cross-sectional SEM images are used to illustrate the realistic structural complexity of printed AuNP assemblies. Accordingly, **Figure 1h–1l** is intended to provide experimental evidence of the dense, disordered, and compositionally mixed nature of the printed assemblies, rather than to serve as a direct structural input for the idealized simulation models discussed below. We therefore use these cross-sectional SEM images for qualitative structural identification rather than for extracting a quantitative coordination-number histogram, because the two-dimensional section does not uniquely determine the three-dimensional particle connectivity in the compositionally mixed assembly.

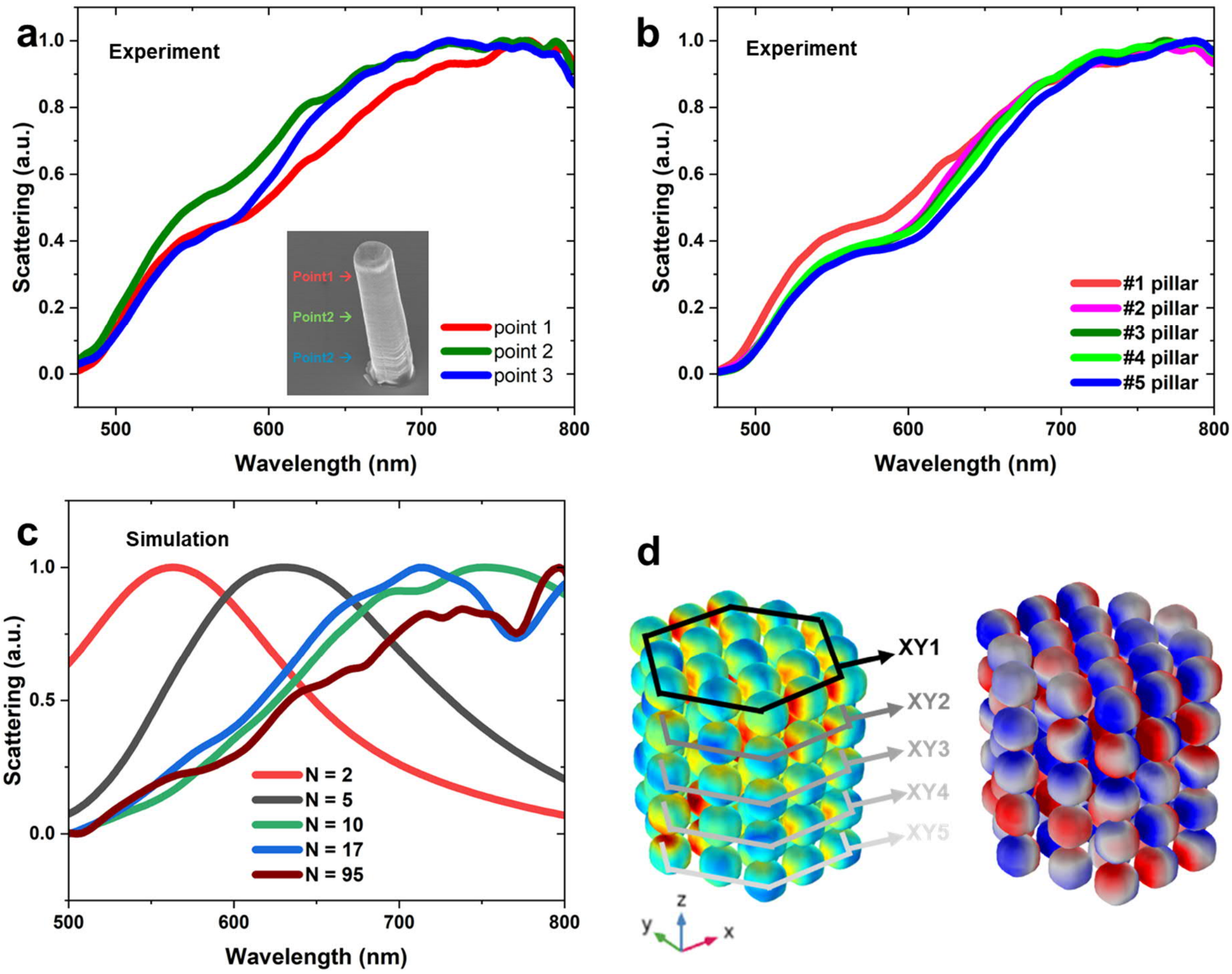

**Figure 2 | (a)** Dark-field scattering spectra acquired from the bottom, middle, and top positions of a single AuNP pillar. **(b)** Dark-field scattering spectra measured from the middle-height regions of five different AuNP pillars labeled #1–#5 in **Figure 1d**. **(c)** Simulated scattering spectra of simplified 3D AuNP assembly models with different numbers of nanoparticles (N = 2, 5, 10, 17, and 95) under a fixed effective interparticle separation. **(d)** Perspective view of simulated near-field intensity (left) and surface charge distribution (right) for a 3D AuNP cluster (N = 95).

Despite this structural disorder, dark-field scattering measurements show that the broad ensemble scattering envelope remains similar when probed at the lower, middle, and upper regions of a single upright pillar (**Fig. 2a**). In these measurements, the printed pillars were kept upright on the substrate and probed under the same top/oblique dark-field geometry using a 100× BF/DF objective. The lower, middle, and upper positions were selected by dividing the pillar height into three regions along the vertical axis, and the spectra were acquired without changing the illumination or collection conditions. Because the printed pillars are large, disordered nanoparticle assemblies, the measured dark-field spectra are interpreted as broad ensemble-averaged scattering envelopes rather than as spectra of discrete, well-resolved oligomeric resonances. In this sense, the measured spectra should be viewed as the result of many damped and spectrally overlapping coupled responses with different radiative strengths, rather than as the selective excitation of a single well-defined plasmonic mode. Therefore, our comparison focuses on the reproducibility of the spectral envelope under identical measurement conditions, rather than on the extraction of individual narrow resonance peaks. This height-wise similarity suggests that the spatially averaged optical response is comparable along the growth direction, consistent with statistically similar packing environments within the probed regions. Moreover, scattering spectra collected from different pillars within the same array show similar broad spectral envelopes, supporting the reproducibility of the ensemble-averaged optical response across nominally independent structures (**Fig. 2b**). The SEM images (**insets of Fig. 1e-1g)** reveal a gradual increase in pillar diameter across the array, which can be attributed to the direct-contact nature of the assembly process using a glass micropipette, where repeated contact with the substrate induces slight nozzle wear over successive fabrication cycles. Importantly, this systematic variation in macroscopic pillar diameter does not lead to discernible changes in the scattering spectra, indicating that the spectral characteristics are largely insensitive to the pillar diameter within the examined range. This observation suggests that the optical response is governed primarily by nanoscale plasmonic interactions within the cluster rather than by the overall pillar geometry. To further elucidate this behavior, finite-difference time-domain (FDTD) simulations were performed using simplified 3D AuNP assembly models with increasing numbers of NPs (N = 2, 5, 10, 17, and 95) and a fixed effective interparticle separation (**Fig. 2c**). The selected particle numbers were not intended to define a unique structural sequence of ideal oligomers, but to sample the evolution from a minimal coupled unit to progressively larger disordered 3D assemblies within a computationally tractable model. In this context, N = 2 provides a dimer-like reference, N = 5 and N = 10 represent small disordered assemblies, N = 17 captures an intermediate regime with appreciable geometry-dependent variation, and N = 95 represents a larger model cluster used to examine the emergence of a stabilized broad scattering envelope. These models are not intended to reproduce the full experimental gap distribution, but to evaluate how the ensemble scattering response evolves with particle number and representative structural perturbations under a fixed geometry. While small-N assemblies exhibit pronounced spectral evolution with increasing N, the simulated scattering spectra for N = 17 and N = 95 show similar broad scattering envelopes under the simplified fixed effective-gap model. To examine the influence of geometrical arrangement, we additionally simulated three independently generated geometries for each of N = 17 and N = 95 (Supporting Fig. S5). These additional calculations show that N = 17 exhibits more noticeable geometry-dependent spectral variations, whereas N = 95 shows a more stable broad scattering envelope with reduced geometry-to-geometry variation. Nevertheless, the broad spectral envelopes of N = 17 and N = 95 remain substantially overlapping within this limited sampling range, indicating that the ensemble-averaged scattering response becomes weakly dependent on the total particle number once the simulated assembly exceeds the relevant local coupling length scale. This size-insensitive behavior is consistent with an ensemble-averaged plasmonic response: once the assembly contains many local coupling units, the far-field spectrum is governed primarily by the statistical distribution of local environments rather than by the exact total particle number or the macroscopic pillar diameter.

Together, the experimental and numerical results in **Figure 2** demonstrate that large 3D AuNP assemblies exhibit spatially uniform and reproducible ensemble optical responses despite variations in macroscopic geometry and intrinsic structural disorder. These findings raise a fundamental question regarding the physical origin of this uniformity: namely, how local electromagnetic interactions within a disordered 3D nanoparticle ensemble give rise to a reproducible, size-insensitive far-field response over the measured pillar dimensions. In the following section, this question is addressed through detailed electromagnetic simulations of near-field intensity and surface charge distributions, providing direct insight into the defect tolerance and averaging mechanisms underlying the observed optical uniformity. To visualize the internal field characteristics of the size-saturated regime, **Figure 2d** presents near-field intensity (left) and surface charge distribution (right) of a representative N = 95 AuNP cluster in a perspective view. The simulated near-field map shows spatially distributed enhancement localized at interparticle junctions throughout the cluster volume, while the corresponding surface charge map reveals a dense ensemble of dipolar charge configurations with varying orientations. Although the geometric model is top-bottom symmetric, the plotted near-field and surface charge distributions represent driven, phase-dependent responses under the selected excitation condition rather than eigenmode symmetry maps. Therefore, the apparent absence of top-bottom symmetry in the visualization mainly reflects the incident-wave direction, retardation across the finite cluster, and the chosen phase reference used for plotting. Together, these results confirm that the optical response is not dominated by a single hotspot or polarization direction, but rather emerges from a volumetric superposition of many local couplings. This perspective-level analysis motivates a more detailed examination of spatial consistency and defect sensitivity across individual XY slices. Here, ensemble-averaging refers to the reduced sensitivity of the broad far-field scattering envelope to individual local perturbations because many local scattering and coupling contributions with different orientations are sampled within the 3D assembly. This should not be interpreted as evidence that all nearest-neighbor couplings are statistically identical.

**Figure 3** builds directly on this volumetric view by analyzing five depth-resolved XY cross-sections of the same cluster. By systematically comparing near-field and surface charge maps under various structural conditions—including defect-free, single-particle vacancy, random voids, and planar removal—this figure reveals how the simulated ensemble near-field distribution responds to different types and scales of structural perturbation. All near-field and surface charge distributions are normalized using a consistent color scale to enable fair comparison across different structural configurations and cross-sectional planes. In the defect-free case, near-field maps reveal many hotspots at interparticle junctions (**Fig. 3a**). These hotspots are widely distributed rather than confined to one region, indicating the optical response is a collective superposition of many local couplings. Importantly, the pattern of hotspots is statistically similar across all five planes, with each slice showing comparable bright and dark regions. This continuity of near-field features across slices indicates that plasmonic interactions are consistent throughout the pillar. The ensemble of many hotspots effectively undergoes volumetric averaging, yielding a uniform near-field profile at the macroscopic scale. Next, a single-particle defect was introduced by removing one NP (**Fig. 3b**). The near-field intensity immediately around the vacancy is slightly reduced, but this change remains local: the rest of each plane's hotspots are essentially unchanged. The overall near-field distribution is qualitatively similar to the intact case. Thus, one missing NP acts as a localized perturbation rather than altering the global response.

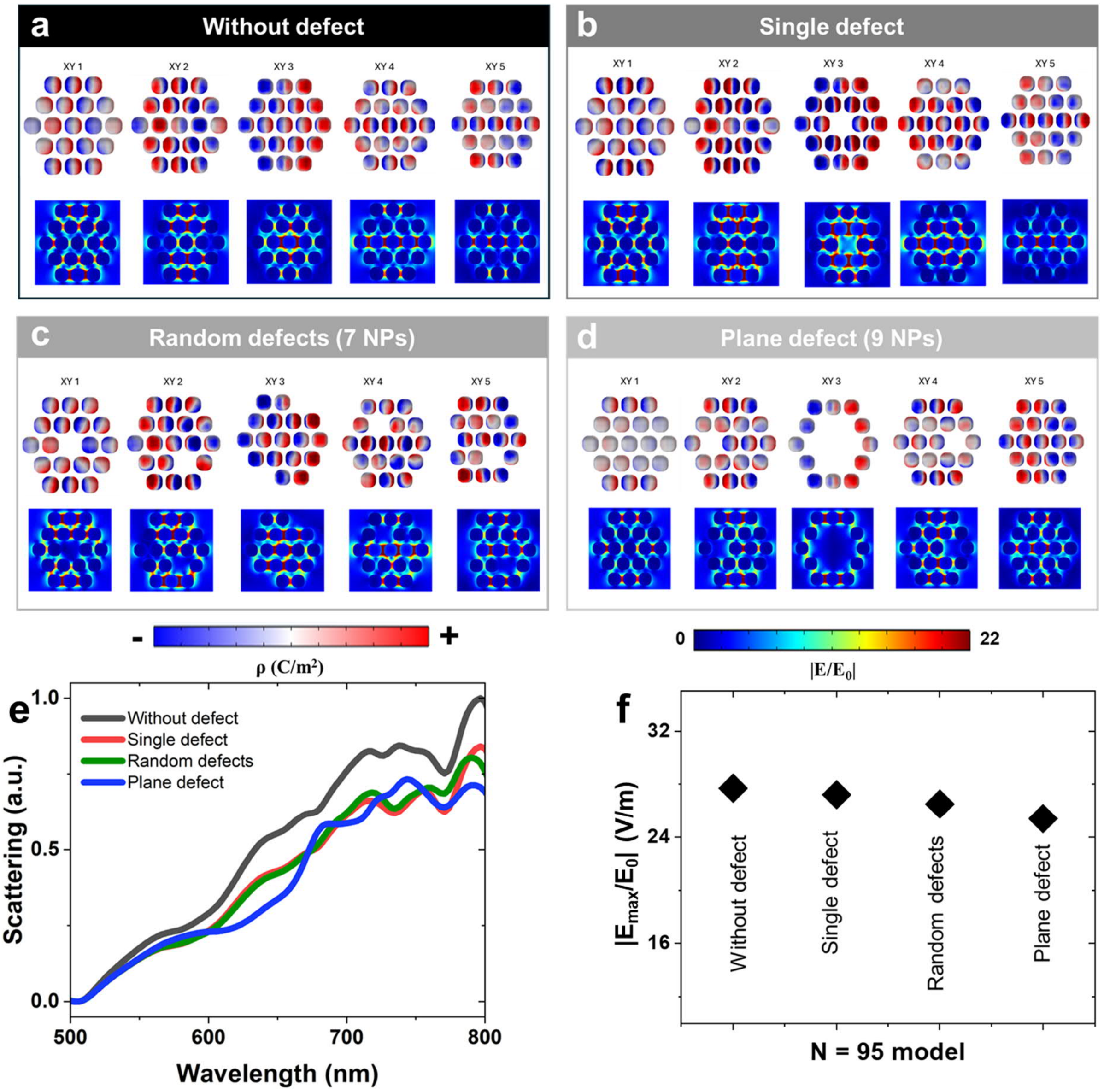

**Figure 3 |** Near-field intensity (bottom) and surface charge (top) distributions of a representative 3D AuNP cluster (N = 95) across five XY planes under different structural conditions: **(a)** defect-free, **(b)** single-particle defect, **(c)** seven randomly distributed defects, and **(d)** planar defect involving nine missing particles. **(e)** Calculated far-field scattering spectra of 3D nanoparticle clusters with and without structural defects. **(f)** Calculated near-field enhancement factor ($|E_{max}/E_0|$) at $\lambda = 770$ nm for 3D AuNP assembly under different defect conditions, showing only marginal reduction in field intensity despite increasing structural perturbations.

In a large 3D cluster, the effect of a single missing particle is distributed across many alternative coupling pathways. As a result, strong enhancements at many junctions remain present despite the local void. For random defects, seven nanoparticles (NPs) were removed. In particular, we ensured that at least one or more NPs are removed from every "XY" plane (**Fig. 3c**). Even under this severe perturbation, the near-field patterns in each XY plane remain remarkably consistent. Hotspots adjacent to the voids shift or diminish, but new hotspots emerge at nearby junctions, so each plane still shows a dense distribution of bright regions. This behavior indicates that the remaining particle network provides alternative coupling pathways around the vacancies. The continuity of near-field patterns across slices indicates that the system's plasmonic modes self-adjust to local voids while preserving the global intensity distribution. Finally, for the plane defect, we removed dominantly seven NPs in "XY 3" plane alongside two NPs in neighboring planes (XY 2 and XY 4) (**Fig. 3d**). This largest perturbation creates a macroscopic void. In the near-field maps, intensity hot spots are lost within the removed plane, but the adjacent planes develop strong hotspots at the edges of the gap. Remarkably, the field enhancements remain broadly distributed in all other regions. Beyond the defect, the hotspot pattern resembles that of the intact cluster, indicating the system effectively reroutes its coupling network around the void. The overall near-field profile remains qualitatively similar to the defect-free case, even in the presence of the planar void. The induced surface charge distributions provide complementary insight. In the defect-free cluster, each NP supports a dipolar charge pattern with random orientation. Introducing any defect causes only local charge rearrangements. Adjacent particles around a vacancy adjust their dipole strength or orientation, yet the overall charge distribution remains largely preserved. With seven random defects, a few particles near voids have perturbed dipoles, but distant particles display the same alternating charge pattern as before. Even with the planar defect, particles at the void edges develop stronger dipoles to maintain continuity. In all cases, this local redistribution of induced charge keeps the collective plasmonic modes essentially intact despite structural imperfections. Additional near-field and surface charge maps taken along the XZ plane further support the defect tolerance observed in the XY-plane analyses (**Supporting Fig. S1**). These orthogonal slices confirm that field continuity and dipolar charge distributions are preserved across the cluster height, even in the presence of single or multi-particle defects. To further elucidate the robustness of plasmonic mode coupling within the 3D nanoparticle assembly, we compare the surface charge density distributions of four representative configurations: defect-free, single defect, random defects, and planar defect – extracted at the XY 3 plane (**Supporting Fig. S2**). Because the surface charge maps represent phase-dependent instantaneous charge distributions under a selected excitation condition, they are used here only as qualitative visualizations of local charge redistribution rather than as a rigorous modal phase analysis. Accordingly, we do not use these maps to assign a unique coupled mode, phase circulation, or polarization-resolved modal symmetry.

Importantly, the spatial phase progression and dipolar orientation remain preserved from an ideal structure to the model study incorporating a large planar defect. Although local charge redistribution occurs in the vicinity of defects, the global oscillation symmetry is maintained. These observations suggest that, within the simplified simulation model, the far-field response can be approximately described by the superposition of many locally coupled dipolar responses rather than by a defect-sensitive mode of a small oligomer. The persistence of this coherent charge distribution across increasing defect severity indicates that the plasmonic excitation is not confined to localized particle–particle junctions, but instead extends volumetrically throughout the structure. Such spatial averaging of many local dipolar responses contributes to the apparent robustness of the ensemble plasmonic response.

Building on the depth-resolved near-field continuity, we next evaluate whether such defect tolerance also persists in the far-field optical response. **Figure 3e** presents the calculated far-field scattering spectra of 3D nanoparticle clusters under various defect conditions. While the defect-free structure exhibits the highest scattering intensity, the overall spectral shape remains nearly unchanged across the single, random, and plane defect cases. This suggests that, despite local disruption of near-field coupling, the collective optical response of the cluster is remarkably tolerant to structural defects. To further quantify defect tolerance, we extracted the near-field enhancement $|E_{max}/E_0|$ at $\lambda$ = 770 nm (**Fig. 3f**). The near-field enhancement reported in **Figure 3f** corresponds to the maximum near-field amplitude normalized to the incident field, $|E_{max}/E_0|$, extracted from the simulated near-field maps at $\lambda$=770 nm for each defect configuration. The introduction of defects results in a limited reduction in the maximum near-field enhancement within the simplified fixed-gap model, even for the planar defect configuration. The present comparison is intended to evaluate relative changes among representative perturbation cases under the same effective-gap condition, rather than to make a direct quantitative comparison with strongly coupled small-gap oligomers. These results support an ensemble-averaging picture in which the broad optical response becomes less sensitive to individual local perturbations because many local coupling contributions are sampled within the 3D assembly. In this picture, the ensemble of locally coupled dipolar responses redistributes field and charge around local perturbations, helping preserve the broad near-field and far-field response within the simplified model. This inherent resilience is a direct consequence of volumetric mode hybridization and extended dipolar coupling, which collectively stabilize the optical response against structural imperfections.

Together, these observations suggest that the optical response of large 3D AuNP assemblies is governed by the ensemble averaging of many local scattering and coupling contributions, rather than by a single precisely defined geometry or individual nanoparticle junction. Through distributed near-field continuity, the system tolerates structural defects by redistributing fields and charges locally. Consequently, once the cluster exceeds a critical size, its scattering spectrum becomes insensitive to additional particles or voids. This provides a direct microscopic explanation for the experimentally observed height-wise uniformity and pillar-to-pillar reproducibility of the scattering spectra (Figure 2), as well as insensitivity to pillar dimensions. The simulated near-field continuity and defect tolerance thus account for why the measured spectra are uniform and highly reproducible across different pillars.

With this defect-tolerant, size-saturated platform established, we next introduce compositional modulation via a core–satellite architecture to deliberately reshape the modal landscape. **Figure 4** demonstrates the controlled formation of 3D core–satellite AuNP assemblies by systematically varying the concentration of the smaller NPs while maintaining an otherwise identical assembly process. In all cases, free-standing pillars with a uniform length of approximately 30 μm are obtained, indicating that the introduction of secondary NPs does not compromise the overall structural integrity or growth stability of the 3D assembly. Low-magnification side-view SEM images **(Fig. 4a, 4d and 4g**) confirm that the macroscopic pillar geometry remains consistent across different experimental conditions, while intermediate-magnification images (**Fig. 4b, 4e and 4h**) reveal that densely packed NP assemblies extend uniformly over the entire pillar surface.

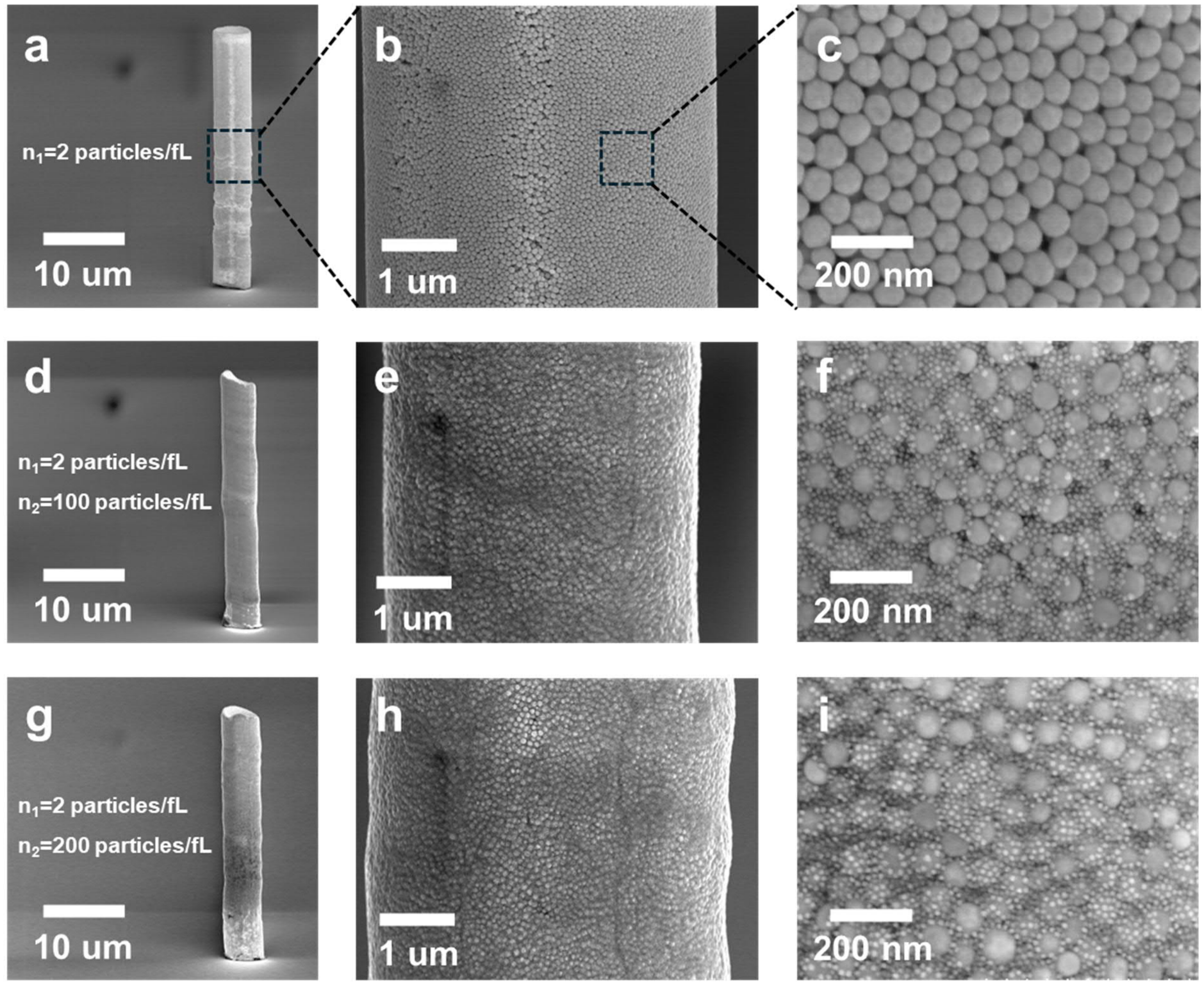

**Figure 4 | (a–c)** SEM images of a core-only AuNP pillar formed with 80 nm AuNPs ($n_1$=2 particles/fL) without 20 nm AuNPs ($n_2$ = 0): **(a)** side-view image showing the overall pillar geometry, **(b)** intermediate-magnification image of the pillar body, and **(c)** high-magnification image showing the local nanoparticle structure. **(d–f)** SEM images of a core–satellite AuNP pillar formed with fixed 80 nm AuNP concentration ($n_1$=2 particles/fL) and moderate 20 nm AuNP concentration ($n_2$=100 particles/fL): **(d)** side-view image, **(e)** intermediate-magnification image, and **(f)** high-magnification image. **(g–i)** SEM images of a core–satellite AuNP pillar formed with fixed 80 nm AuNP concentration ($n_1$=2 particles/fL) and higher 20 nm AuNP concentration ($n_2$=200 particles/fL): **(g)** side-view image, **(h)** intermediate-magnification image, and **(i)** high-magnification image.

These observations establish that compositional modulation can be introduced without disrupting the robust, large-scale architecture of the 3D AuNP assemblies. High-magnification SEM images (**Fig. 4c, 4f and 4i**) provide direct insight into the evolution of the nanoscale architecture as the concentration of 20 nm AuNPs ($n_2$) is increased relative to the fixed concentration of 80 nm AuNP cores ($n_1$). Without $n_2$, the surface is dominated by closely packed 80 nm particles, whereas increasing $n_2$ leads to progressive incorporation of the smaller NPs, resulting in a pronounced core–satellite morphology. Notably, the 20 nm AuNPs do not merely decorate the outer surface of the pillar but are incorporated throughout the interparticle gaps within the assembly (**Supporting Fig. S3**). This internal distribution, which has been independently verified through cross-sectional observations, indicates that the satellite NPs permeate the 3D network and populate the nanoscale voids between neighboring cores in a manner comparable to their surface coverage. Such volumetric incorporation is consistent with the meniscus-mediated assembly process, in which NPs are continuously supplied and immobilized within a confined liquid bridge, allowing smaller particles to access and occupy available interstitial regions during pillar growth. This volumetric incorporation provides a structural basis for composition-dependent modulation of local coupling pathways throughout the 3D assembly. Therefore, the core–satellite architecture serves not only as a structural demonstration of multiscale nanoparticle assembly, but also as a composition-tunable 3D plasmonic scaffold for wavelength-dependent near-field and SERS responses.

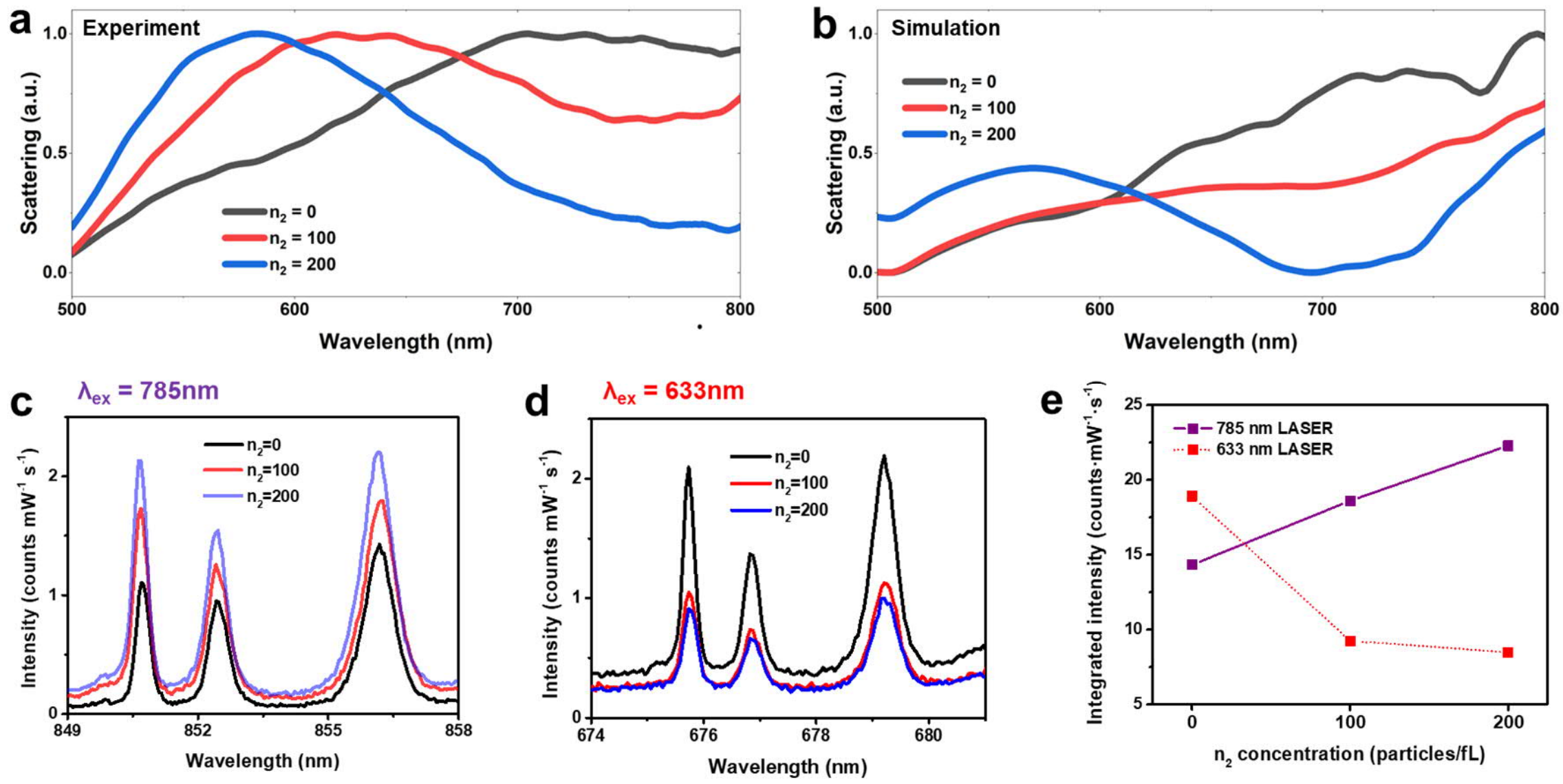

**Figure 5 | (a)** Dark-field scattering spectra of 3D AuNP assemblies with different compositions, comparing a core-only assembly and core–satellite assemblies formed with increasing concentrations of 20 nm AuNPs. **(b)** Simulated far-field scattering spectra of the N = 95 model with varying numbers of satellite nanoparticles. **(c)** Power- and time-normalized wavelength-resolved SERS spectra of benzenethiol measured under 785 nm excitation for AuNP assemblies with different satellite concentrations. **(d)** Power- and time-normalized wavelength-resolved SERS spectra of benzenethiol measured under 633 nm excitation for AuNP assemblies with different satellite concentrations. The selected spectral windows in (c,d) are used to clearly resolve representative benzenethiol Raman bands relevant to the composition-dependent comparison. **(e)** Power- and time-normalized integrated SERS

intensity of the benzenethiol C–S stretching band as a function of satellite concentration under 633 nm and 785 nm excitation.

**Figure 5** examines how compositional modulation of the 3D AuNP assemblies, realized through the core–satellite architecture in Figure 4, influences their plasmonic response and its interaction with optical excitation. The incorporation of 20 nm AuNPs modifies the optical response through the heterogeneous coupling environment of the printed core–satellite assembly. The smaller AuNPs can perturb direct coupling pathways among neighboring 80 nm AuNPs by occupying interstitial regions, which may contribute to the blueshift of the broad scattering envelope observed in **Figure 5a**. At the same time, because the 20 nm AuNPs are plasmonically active, they can participate in local core–satellite near-field coupling and introduce additional nanoscale junctions. Thus, the spectral evolution is attributed to a composition-dependent redistribution of local coupling pathways within the 3D plasmonic nanoparticle network, rather than to a simple increase or decrease in the overall coupling strength. This composition-dependent spectral redistribution provides a useful platform for probing wavelength-dependent optical behavior within a structurally uniform 3D pillar. To further examine this compositional trend, we performed simulations using the N = 95 assembly model populated with different numbers of smaller satellite AuNPs. The satellite AuNPs were distributed around the larger particles using a Gaussian distribution approach, with their diameters varied between 16 and 20 nm to mimic the experimental satellite population (**Supporting Fig. S4**). Although this model does not reproduce the full structural complexity of the experimental core–satellite assemblies, it captures the qualitative spectral trend observed experimentally. As the number of satellite AuNPs increases, the long-wavelength scattering contribution is progressively suppressed, while the shorter-wavelength contribution becomes more pronounced (**Fig. 5b**). This trend supports the interpretation that satellite incorporation reshapes the coupling landscape of the larger-particle network and redistributes the ensemble scattering response. The remaining differences between the simulated and experimental spectral positions are likely due to the simplified satellite distribution and the limited structural complexity of the model compared with the experimentally printed core–satellite assemblies.

To explore how the composition-dependent spectral redistribution affects optical excitation, SERS spectra of benzenethiol were measured under two representative excitation wavelengths, 633 nm and 785 nm. In this work, SERS is used as a spectroscopic probe of wavelength-dependent near-field response rather than as a quantitative sensing metric. Notably, a recent study using a SERS platform fabricated by a similar meniscus-guided 3D nanoprinting process reported good spatial SERS uniformity along the pillar height, supporting the practical feasibility of such 3D printed plasmonic assemblies as SERS-active structures [52]. Accordingly, the SERS spectra are presented as a function of wavelength instead of the conventional Raman shift to facilitate comparison with the broad scattering envelope and the excitation wavelengths.

A notable trend emerges when the excitation wavelength is considered together with the satellite concentration. Under 785 nm excitation, which overlaps with the longer-wavelength spectral region affected by satellite incorporation, the Raman signal increases with increasing satellite concentration (**Fig. 5c**). This behavior suggests that the incorporated 20 nm AuNPs introduce additional local junctions and near-field-active coupling pathways that enhance Raman excitation in this spectral region. In contrast, under 633 nm excitation, the Raman signal decreases as the satellite concentration increases (**Fig. 5d**), consistent with the redistribution of the ensemble optical response away from the core-dominated, shorter-wavelength scattering contribution. For clarity, the SERS spectra in **Figure 5c,d** are displayed over selected wavelength windows that clearly resolve representative benzenethiol Raman bands relevant to the composition-dependent comparison. Full-range SERS spectra of the $n_2 = 0$ sample, including four major benzenethiol peaks under both 633 and 785 nm excitation, are provided in **Supporting Figure S6**. This wavelength-dependent trend supports the view that satellite incorporation redistributes local plasmonic coupling pathways within the heterogeneous 3D assembly. The integrated values further support this excitation-dependent behavior (**Fig. 5e**). The integrated SERS intensity was extracted from the representative benzenethiol Raman band assigned to the C–S stretching mode (~1076 $cm^{-1}$), corresponding to scattered wavelengths of ~679 nm and ~857 nm under 633 nm and 785 nm excitation, respectively. For each spectrum, the SERS signal was quantified by integrating the peak area within a ±15 $cm^{-1}$ window after linear baseline subtraction. To enable a more direct comparison between the two excitation wavelengths, all SERS spectra and integrated peak areas in **Figure 5c–5e** were normalized by the estimated laser power and exposure time and are reported as counts $mW^{-1}$ $s^{-1}$. This integration-based analysis provides a robust metric for comparing Raman responses across different excitation and compositional conditions.

Together, the results in **Figure 5** demonstrate that compositional engineering within a structurally uniform 3D AuNP assembly provides a practical route to tune the ensemble optical response by redistributing local plasmonic coupling pathways. By incorporating plasmonically active satellite AuNPs into the 80 nm AuNP network, the core–satellite architecture enables wavelength-dependent modulation of far-field scattering and near-field-enhanced Raman responses without disrupting the macroscopic pillar geometry.

## 3. **Conclusion**

In summary, we have demonstrated that large 3D AuNP assemblies fabricated by meniscus-guided assembly exhibit highly uniform, reproducible, and defect-tolerant ensemble plasmonic responses despite intrinsic structural disorder. Spatially resolved optical measurements show that scattering spectra remain invariant along the height of individual pillars and across different structures, while numerical simulations reveal that this robustness arises from an ensemble averaging over many local coupling regions within the printed nanoparticle assembly. Simulations based on a simplified fixed-gap model suggest that localized defects mainly perturb the electromagnetic response in their immediate vicinity, while the broad ensemble optical response remains comparatively stable. Leveraging this intrinsically robust platform, we further introduced compositional modulation through a core–satellite architecture, enabling the local coupling environment to be tuned without compromising the macroscopic pillar geometry. The incorporated 20 nm AuNPs act as plasmonically active structural modifiers: they can perturb direct coupling pathways among neighboring 80 nm AuNPs while also introducing additional local core–satellite junctions. Wavelength-dependent Raman measurements show contrasting responses under 633 and 785 nm excitation, indicating that satellite incorporation redistributes local plasmonic coupling pathways and near-field-enhanced optical responses within the heterogeneous 3D assembly. Although meniscus-guided 3D printing is a serial direct-writing process with limited throughput compared with wafer-scale lithography or nanoimprint lithography, it provides a deterministic prototyping platform for fabricating freestanding 3D nanoparticle assemblies with controlled geometry and composition. Thus, the present work focuses on mechanistic understanding and design-rule extraction rather than large-area mass production. Future parallelization using multi-nozzle printing or pre-patterned assembly sites may improve its scalability for practical applications. These findings establish a connection between structural statistics, ensemble plasmonic behavior, and composition-dependent spectral tuning in large 3D nanoparticle assemblies, offering a general framework for designing complex plasmonic architectures that combine robustness with tunable optical functionality.

## 4. **Experimental Section/Methods**

***Preparation of the ink:*** PVP-coated 80 nm AuNPs dispersed in water (5 mg/mL) and PVP-

coated 20 nm AuNPs dispersed in water (5 mg/mL) were used as the primary building blocks in the self-assembly process (Nanocomposix, San Diego, CA 92111, USA). PVP-coated AuNPs were selected because the PVP layer provides steric stabilization in aqueous dispersion, suppresses uncontrolled aggregation during evaporation-driven meniscus-guided assembly, and helps maintain printable colloidal inks. According to the supplier-provided specifications, the AuNPs had nominal diameters of 80 nm and 20 nm, with a coefficient of variation (CV) of less than 25%. Therefore, the particles should be regarded as nominally size-defined but not perfectly monodisperse. The exact PVP shell thickness was not independently measured in this study. Nevertheless, the PVP coating is expected to contribute to nanoscale interparticle separation in the printed assemblies, together with finite particle dispersion and local packing disorder. All solutions used in the experiments were purified by repeated centrifugation and washing with deionized (DI) water to remove residual byproducts from the synthesis process prior to use.

***Meniscus-guided 3D nanoprinting of the NPs:*** A custom-built three-dimensional (3D) printing setup was employed, comprising three principal components: (i) an x–y–z motorized translation stage (M-VP-25XA-XYZR, Newport, Irvine, CA, USA) controlled by an XPS-D4 motion controller (Newport), (ii) an x–y–z manual positioning stage (PT3A/M, Thorlabs, Newton, NJ, USA), and (iii) an optical observation system. The optical system consisted of a halogen fiber illuminator (OSL2, Thorlabs), an objective lens (MY20X-804, Mitutoyo, Kanagawa, Japan), and a charge-coupled device (CCD) camera for real-time monitoring of the printing process. Glass micropipettes were fabricated using a commercial pipette puller (P-97, Sutter Instruments, Novato, CA, USA). During printing, the substrate was mounted on the x–y–z motorized stage, while the micropipette was fixed to a custom-designed holder attached to the x–y–z manual stage. NP dispersions were loaded into the micropipette using a syringe. The micropipette position was aligned within the field of view of the CCD camera by adjusting the manual stage, enabling direct visualization of the meniscus formation and growth process. Contact between the micropipette and the substrate, as well as subsequent printing, was controlled exclusively through the motorized stage. The vertical withdrawal of the micropipette during pillar growth was carefully regulated at a constant speed of 300 nm/sec. Detachment was performed by retracting the micropipette at a speed of 25 mm/sec.

***Optical measurements of the plasmonic nanostructures:*** Dark-field scattering measurements were performed using a bright-field/dark-field optical microscope (BX53M, Olympus, Tokyo, Japan). Broadband illumination was provided by a halogen light source (MR16, Philips, Amsterdam, Netherlands), and top/oblique dark-field illumination was implemented using a dedicated BF/DF objective. The printed AuNP pillars were measured in their as-fabricated upright configuration on the substrate. Spatially resolved spectra in Figure 2 were acquired using a 100× BF/DF objective lens with NA = 0.8 and WD = 3.3 mm (LMPLFLN100XBD, Olympus). The bottom, middle, and top positions were defined by dividing the pillar height into three regions along the vertical pillar axis, and the collection spot was positioned at the lower, middle, and upper regions using the optical microscope image. Scattered light was collected through the same objective and delivered to a fiber-coupled spectrometer (USB2000+, Ocean Insight, Orlando, FL, USA) using an optical fiber with a 50 μm core diameter, corresponding to an estimated collection area of approximately 0.5 μm at the sample plane under 100× magnification. All position-dependent and pillar-to-pillar spectra in Figure 2 were acquired under identical illumination, collection, exposure, and acquisition conditions. To reduce spectral noise, the acquired scattering spectra were smoothed using a Savitzky–Golay filtering algorithm implemented in the data analysis software (OriginPro 9.0). Surface-enhanced Raman scattering (SERS) measurements were carried out using a commercial Raman spectrometer (Horiba, LabRAM HR-800, CNU center for Research facilities, Daejeon, Korea). Benzenethiol SERS spectra were acquired using 633 nm and 785 nm excitation lasers. For both excitation wavelengths, the nominal laser power was estimated to be approximately 40 mW based on the 50 mW laser output and the OD-based ND filter setting. Each spectrum was acquired with an exposure time of 10 s and two accumulations. Because the exported spectra correspond to the averaged signal over the two accumulations, the SERS spectra in Figure 5c,d were normalized by the estimated laser power and the 10 s exposure time and are reported as counts $mW^{-1}$ $s^{-1}$. The diffraction-limited laser spot diameters were estimated to be approximately 1.5 μm and 1.9 μm for 633 nm and 785 nm excitation, respectively, based on the objective NA of 0.5.

***Optical simulation of plasmonic nanoclusters:*** A three-dimensional electromagnetic Maxwell solver (COMSOL Multiphysics 6.1, Wave Optics module) package is utilized to simulate near-field amplitude, and surface charge density distribution mappings of plasmonic NP cluster model(s).[53-55] To mimic the experimental shape of bigger NPs (diameter of 80 nm), we used a cube geometry and applied fillet rounding condition on all its edges. The smaller satellite NPs are modeled as spheres with diameters in the range 15–20 nm. In the simulation model, the interparticle separation between neighboring 80 nm AuNPs was set to 20 nm as an effective representative separation. This value should not be interpreted as a directly measured or uniform experimental gap, because the printed assemblies contain closely packed junctions with apparent nanoscale separations, including regions that appear to be below approximately 10 nm in high-resolution SEM images(Fig. 4c), as well as wider local gaps arising from particle-size dispersion, PVP coating, stochastic packing, and local void-like irregularities. The model was therefore used to compare relative trends in the ensemble optical response under a fixed, computationally tractable geometry rather than to reproduce the full experimental gap distribution. For the geometry-dependence analysis in Supporting Figure S5, three independently generated geometries were simulated for each of N = 17 and N = 95 while keeping the particle number, effective interparticle separation, material parameters, and optical boundary conditions identical. The NP cluster is embedded in a dielectric environment with refractive index set to n = 1.5 (approximation made for PVP coated Au NP). Outside the NP cluster, the refractive index is set to air. The optical constants of gold are taken from the Johnson and Christy database.[56] A Gaussian random distribution function is applied to populate the satellite NPs in N = 95 plasmonic cluster model, enabling systematic variation of NP number and the minimum–maximum diameter range. For the scattering simulations, a broadband plane-wave excitation was used as the source to obtain spectra over the simulated wavelength range. The same excitation configuration was used for all cluster models to enable direct comparison. The plasmonic scattering spectra were extracted using a three-dimensional box-shaped power monitor positioned close to the nanoparticle cluster in ANSYS Lumerical FDTD. The near-field amplitude profiles and three-dimensional surface charge density distributions were calculated using COMSOL Multiphysics. Two dimensional slices in XZ and XY directions are applied to calculate the cross-sectional amplitude near-field amplitude profiles. Three-dimensional surface charge density distribution mappings are calculated by considering the outward normal vector, skin effect and local electric field.[53-55] A general mesh size of 10 nm is used, with a mesh override of 1 nm in the vicinity of the NP cluster to ensure high accuracy. The spectral ranges used for the scattering measurements and simulations were chosen to cover the reliable visible-to-near-infrared detection window of the microscope–spectrometer system and the spectral region relevant to the 633 and 785 nm excitation conditions, including the corresponding Raman-scattered wavelengths analyzed in Figure 5. Although additional long-wavelength spectral contributions may extend beyond this range, the present analysis focuses on the common spectral window in which the experimental scattering spectra, simulated scattering trends, and wavelength-resolved SERS responses can be directly compared.

**Acknowledgements**

This research was supported by NRF Sejong Science fellowship (NRF-2021R1C1C2011447) and the regional innovation system & education (RISE)-(Specialized Industry Scale-up) program through the Gyeongbuk RISE Center, funded by the Ministry of Education (MOE) and the Gyeongsangbuk-do, Republic of Korea (2025-rise-15-105). This research was also supported by the Basic Science Research Program through the National Research Foundation of Korea (NRF) (grant no. RS-2023-00219703). The authors acknowledge the financial support from the German Ministry of Education and Research (BMBF) within the PhoQuant project (grant number 13N16103).

**Author Contributions**

V.D. and W.-G.K conceptualized and developed the work. S.K. prepared the samples. S.K. and V.D. performed the optical measurements and SEM characterization. V.D. performed optical simulations. S.K., H.K. and W.-G.K assembled experimental set-up for meniscus-guided 3D nanoprinting. V.D., S.K., W.-G.K. and J.M.L. drafted the manuscript. T.Z. provided important revision. S.-K.S. provided valuable feedback during the revision process. T.Z., J.M.L. and W.-G.K. supervised and managed this work. V.D. and S.K contributed equally to this work.

Supporting Information

**Spatially Uniform and Defect-Tolerant Plasmonic Responses in 3D printed Gold Nanoparticle Assemblies**

*Vasanthan Devaraj, Sunghyun Kwak, Hyeongjip Kim, Sang-Keun Sung, Jong-Min Lee*, Thomas Zentgraf * and Won-Geun Kim**

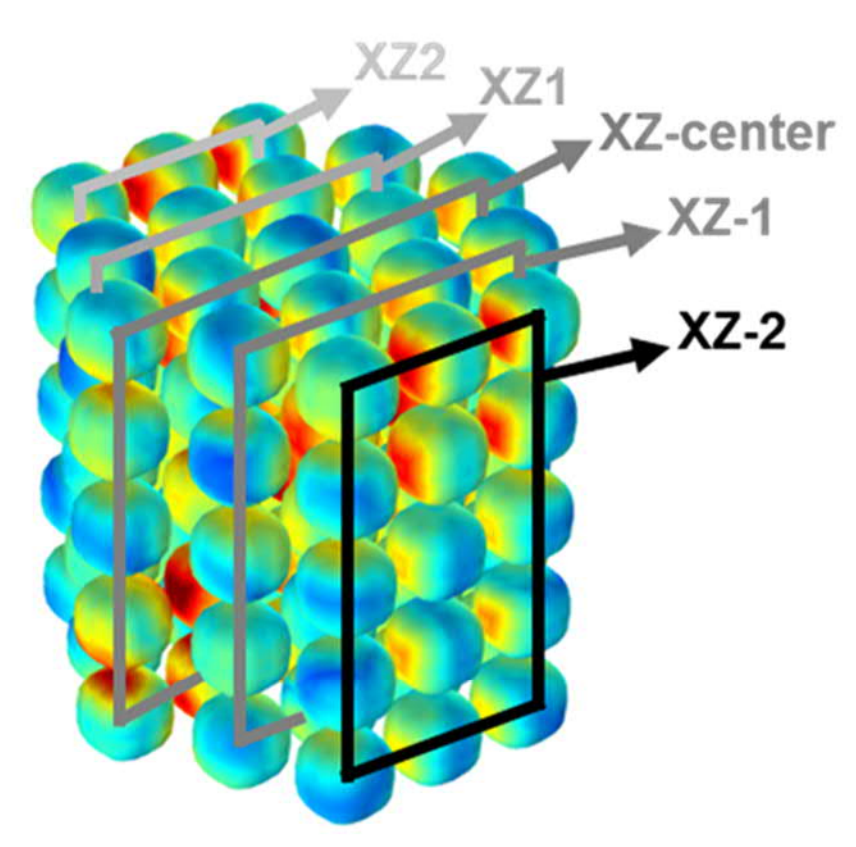

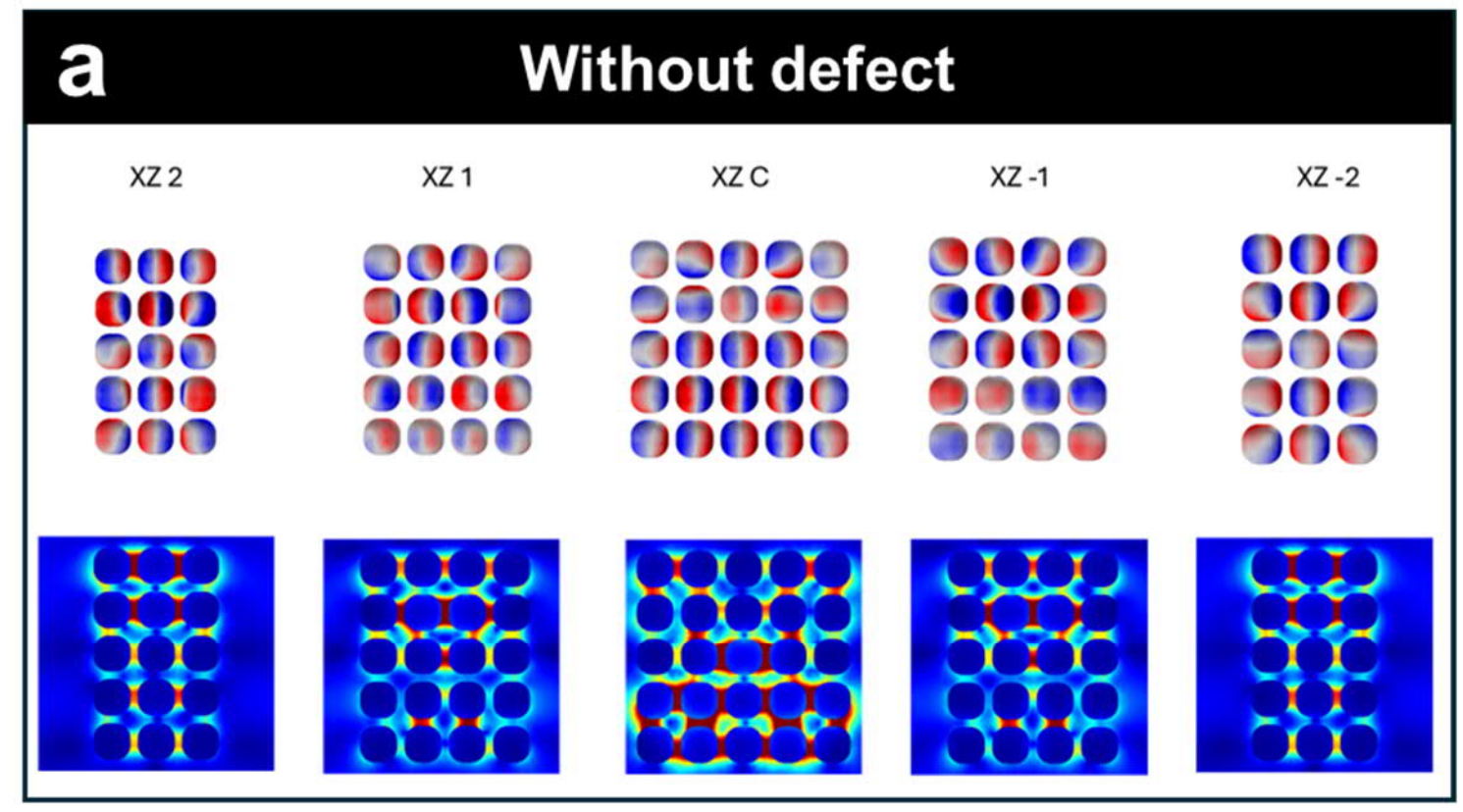

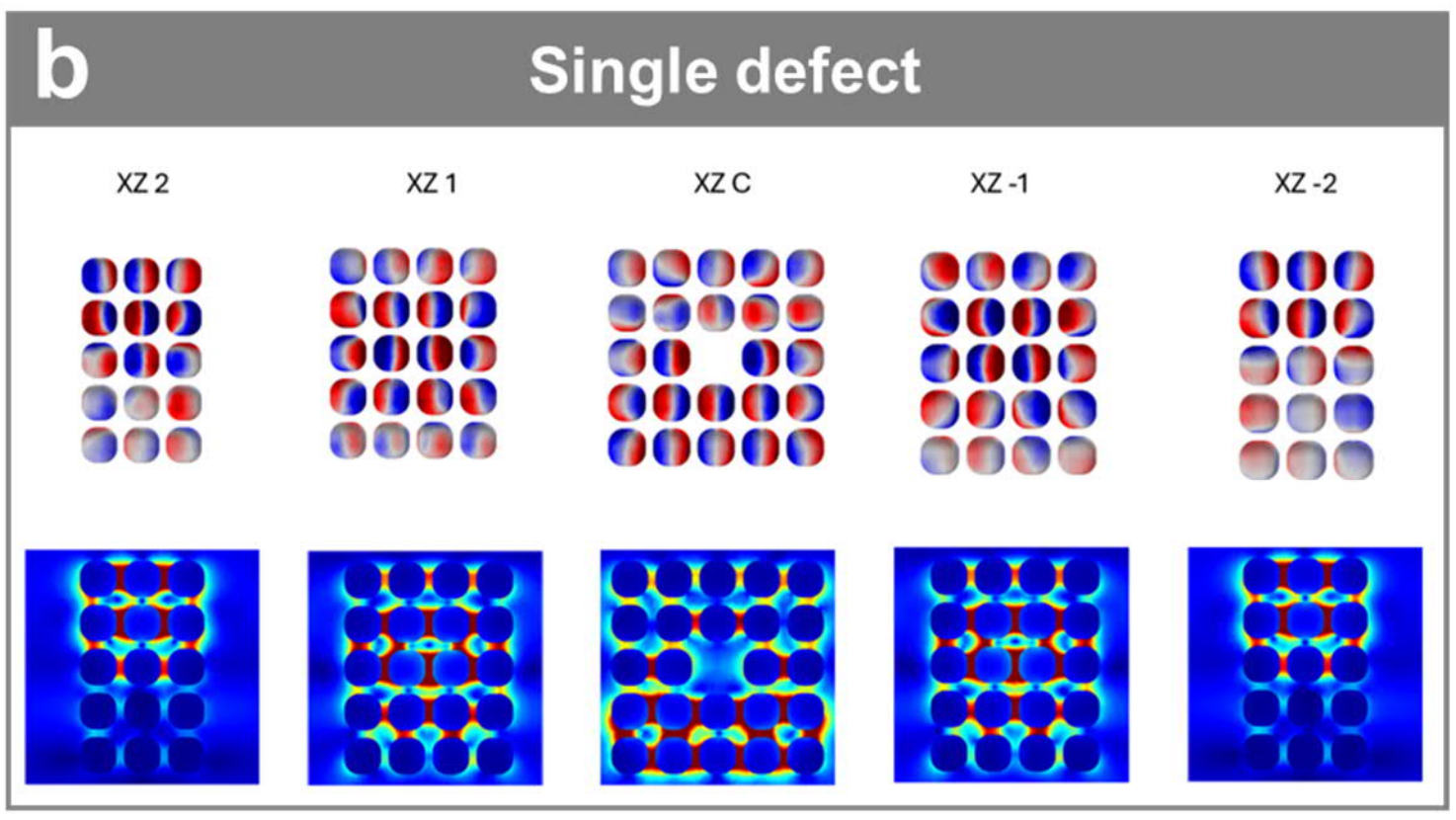

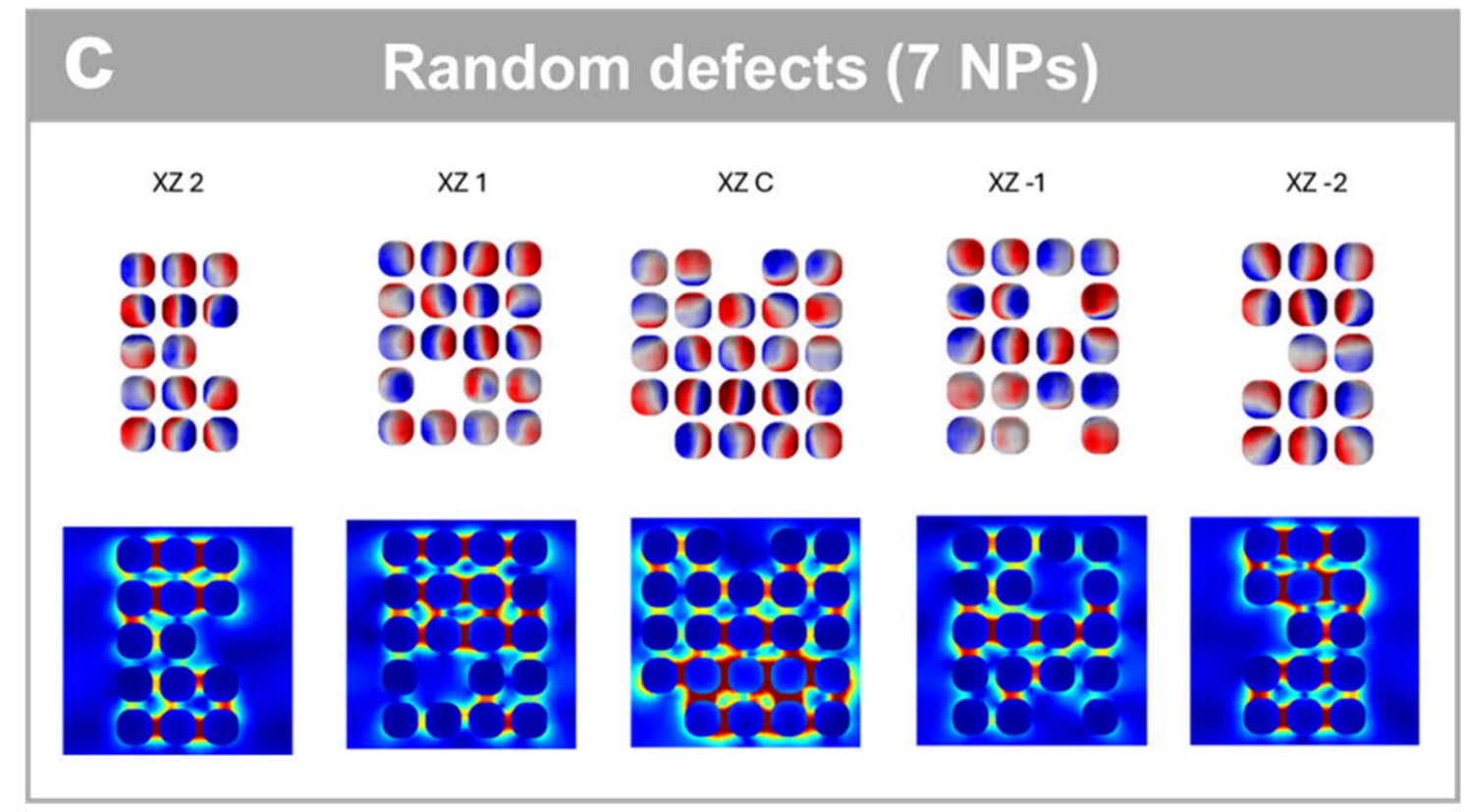

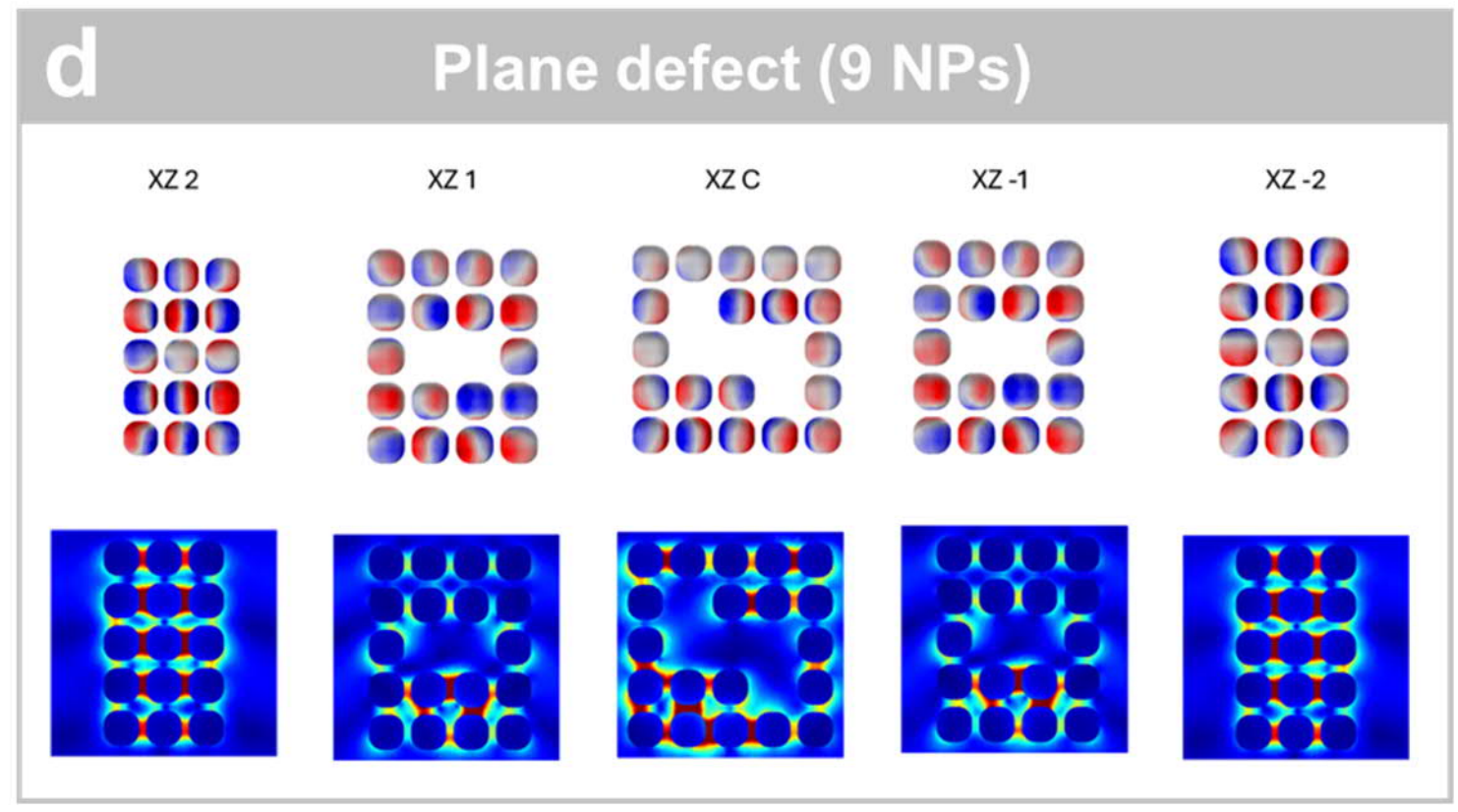

**Supporting Figure S1.** Cross-sectional near-field intensity (bottom row) and surface charge (top row) distributions taken along the XZ plane for a representative 3D AuNP cluster (N = 95) under different structural conditions: (a) defect-free, (b) single-particle defect, (c) seven

randomly distributed defects, and (d) planar defect involving nine missing nanoparticles. Across all cases, the near-field enhancements remain broadly distributed throughout the cluster height, with only localized perturbations near defect regions. The corresponding surface charge maps reveal minor local dipole reconfigurations while preserving the overall collective charge distribution. These XZ-plane analyses corroborate the volumetric averaging behavior observed in the XY-plane slices, confirming the three-dimensional defect tolerance of the plasmonic modes.

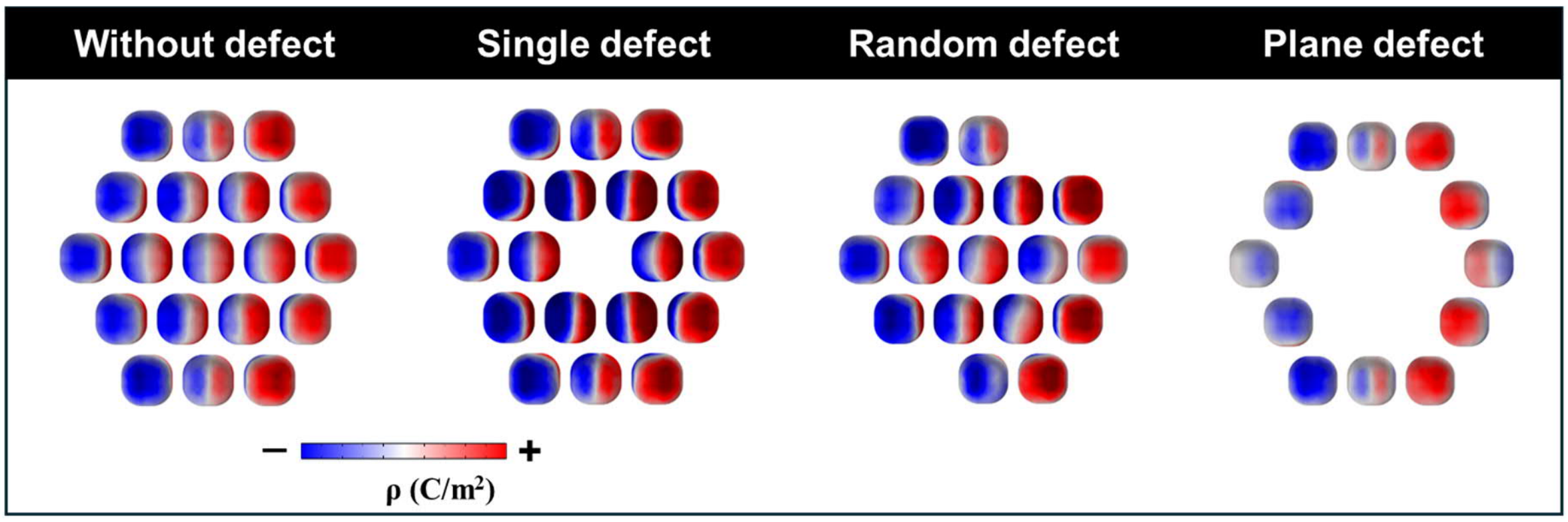

**Supplementary Figure S2.** Surface charge density distributions extracted at the XY3 plane for four representative configurations: defect-free, single defect, random defects, and planar defect. The maps are shown as representative phase-dependent snapshots under the selected excitation condition and are used to visualize qualitative local induced-charge redistribution around defect regions. Because the apparent charge pattern can depend on the excitation phase, retardation, and the chosen phase reference used for visualization, the maps are treated here as qualitative charge-redistribution snapshots rather than as a phase-resolved modal analysis or evidence of a unique phase circulation. The comparison illustrates that representative defects mainly induce local charge redistribution within the simplified simulation model.

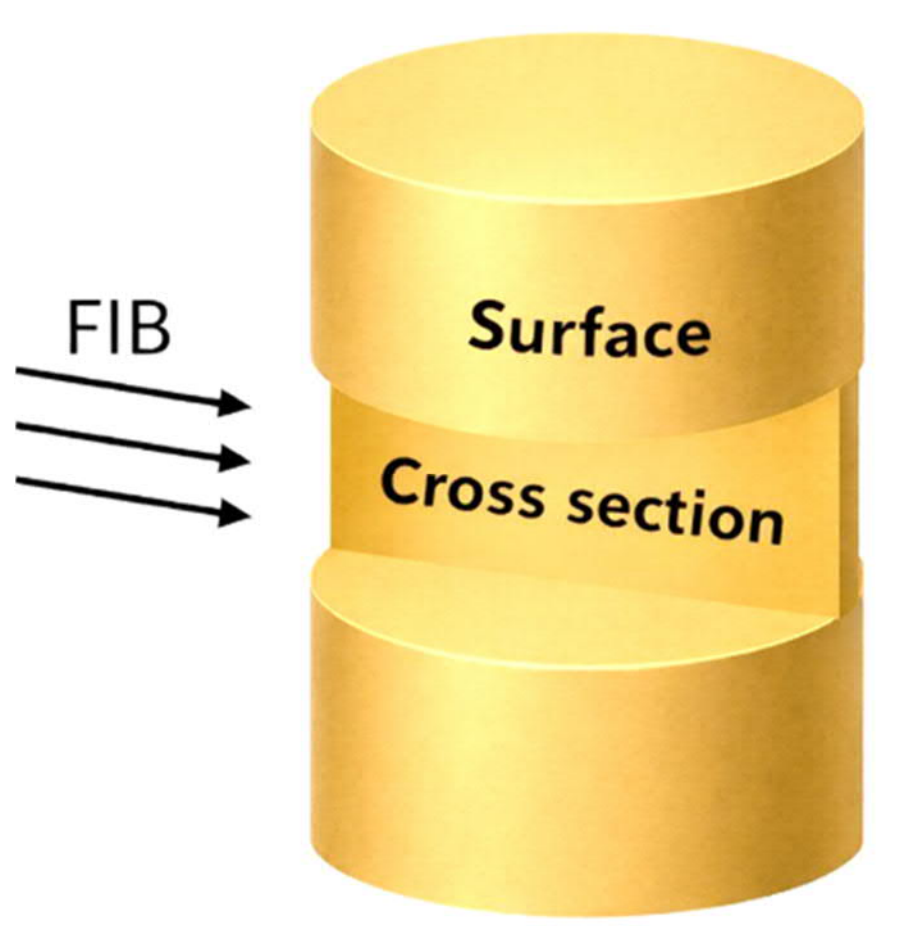

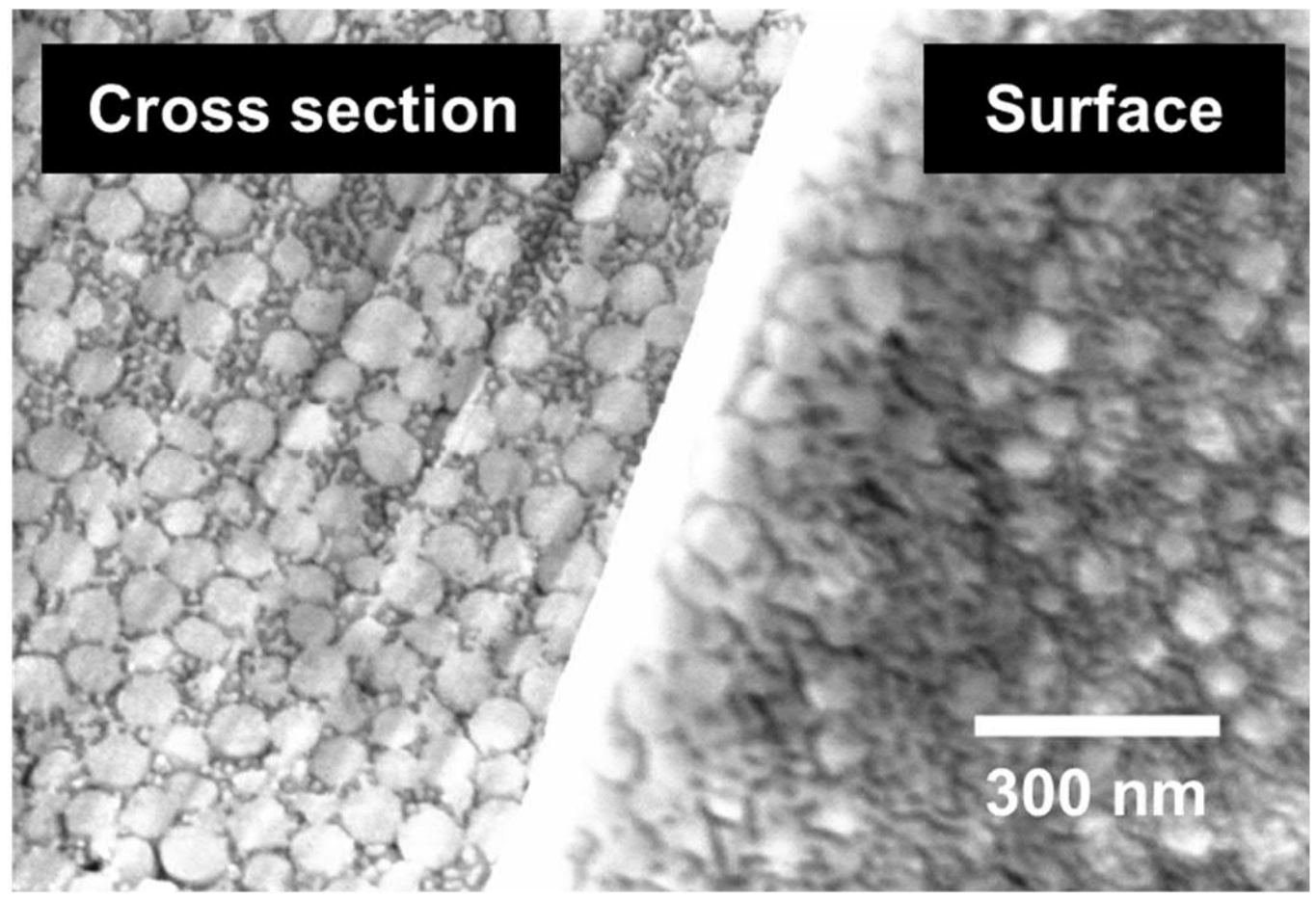

**Supplementary Figure S3.** Focused ion beam (FIB) milling and SEM images showing the cross-sectional and surface morphology of the core–satellite AuNP pillar. The cross-sectional view confirms that the 20 nm satellite AuNPs are distributed throughout the internal interparticle gaps of the 3D cluster, rather than being confined solely to the outer surface (scale bar: 300 nm).

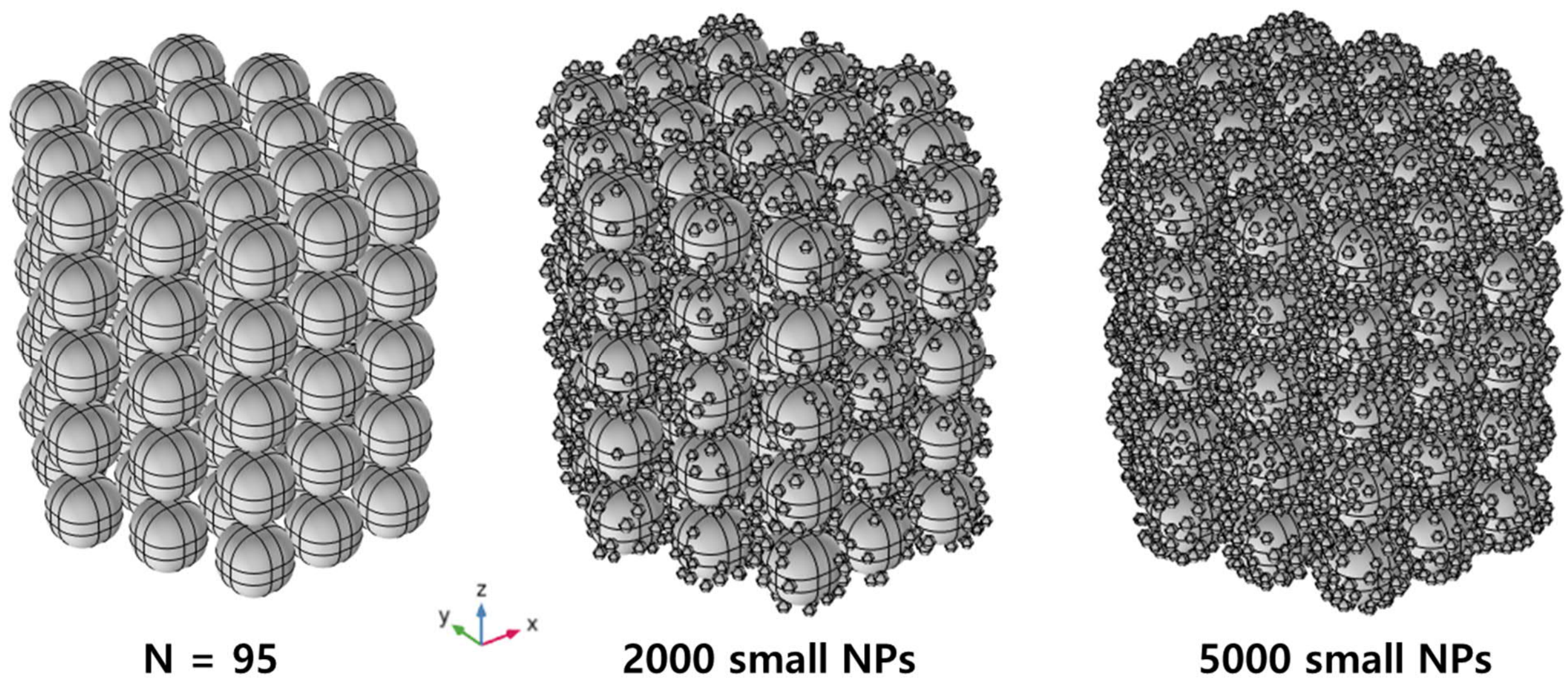

**Supporting Figure S4.** Three-dimensional COMSOL models of the N = 95 core cluster populated with satellite nanoparticles following a Gaussian size distribution (16–20 nm diameter). Representative configurations of N = 95 model and an introduction of 2000 and 5000 smaller AuNPs which illustrates the progressive volumetric filling of interstitial regions as the satellite concentration increases. These models were used to simulate the compositional dependence of the plasmonic scattering spectra shown in Figure 5.

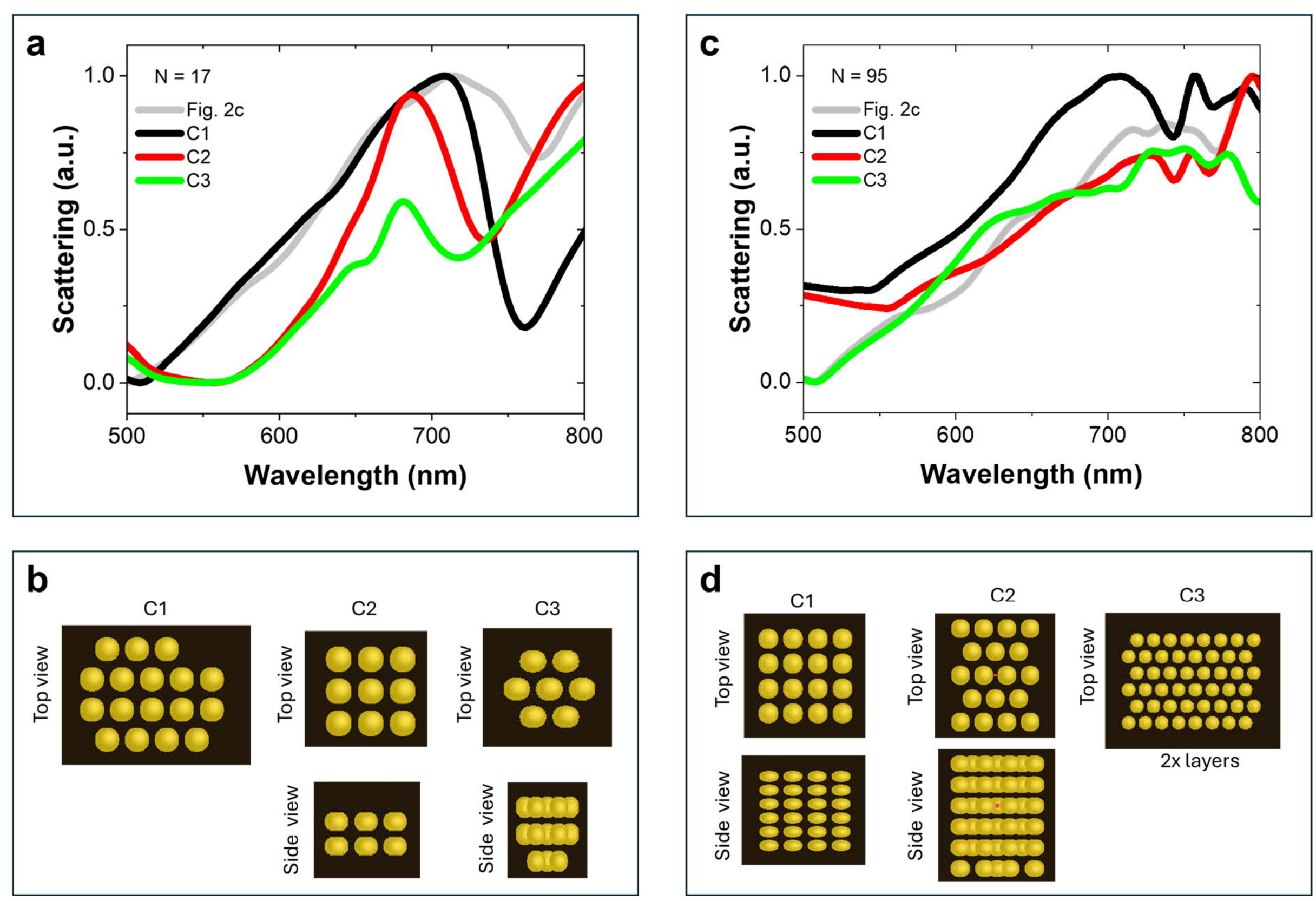

**Supporting Figure S5.** Geometry-dependent scattering spectra and corresponding simulation models of simplified N = 17 and N = 95 AuNP assemblies. (a) Simulated scattering spectra of three independently generated N = 17 geometries, together with the original N = 17 model used in Figure 2c. (b) Top- and side-view images of the three N = 17 simulation models used for the geometry-dependence analysis. (c) Simulated scattering spectra of three independently generated N = 95 geometries, together with the original N = 95 model used in Figure 2c. (d) Top- and side-view images of the three N = 95 simulation models. All geometries were simulated under the same fixed effective interparticle separation and optical conditions used in the main simulations. The total number of AuNPs was kept constant for each set, while the spatial arrangement of the particles was varied to examine geometry-dependent spectral variation. The N = 17 models show noticeable variation in scattering intensity and spectral profile, indicating that smaller assemblies remain sensitive to specific nanoparticle arrangements. In contrast, the N = 95 models exhibit a more stable broad scattering envelope with reduced geometry-to-geometry variation, consistent with stronger averaging over many local coupling regions. Although this sampling is not statistically exhaustive, the comparison supports the qualitative interpretation that geometry-dependent spectral variation is reduced as the assembly size increases within the simplified fixed effective-gap model.

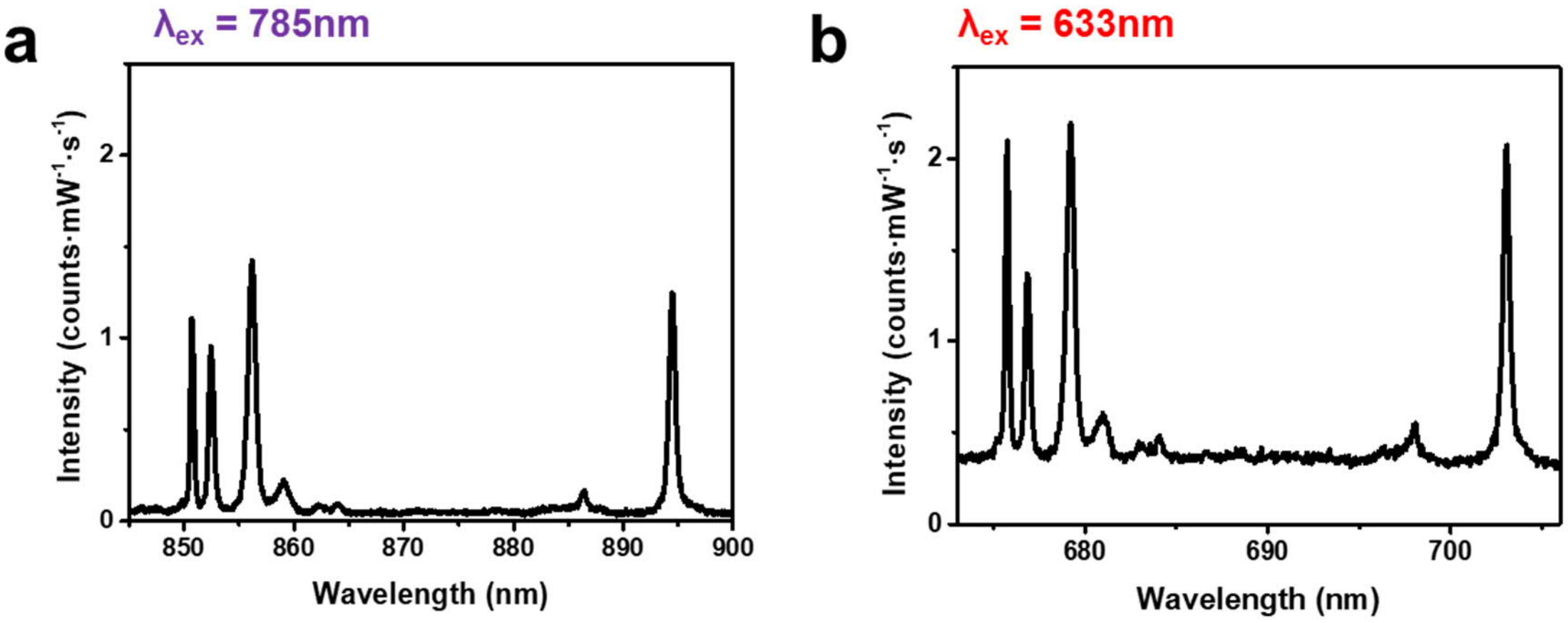

**Supporting Figure S6.** Full-range SERS spectra of benzenethiol measured from the $n_2 = 0$ AuNP assembly under (a) 785 nm and (b) 633 nm excitation. The wider spectral windows include four major benzenethiol Raman bands, confirming the characteristic benzenethiol spectral signature under both excitation conditions. These full-range spectra complement the selected spectral windows shown in Figure 5c,d, which are used to improve the visibility of representative Raman bands and composition-dependent intensity trends.